\documentstyle[12pt]{article}
\begin{document}
\def\om{\omega}
\def\omt{\tilde{\omega}}
\def\ti{\tilde}
\def\o{\Omega}
\def\bchi{\bar\chi^i}
\def\In{{\rm Int}}
\def\ba{\bar a}
\def\w{\wedge}
\def\ep{\epsilon}
\def\k{\kappa}
\def\Tr{{\rm Tr}}
\def\ST{{\rm STr}}
\def\ss{\subset}
\def\ot{\otimes}
\def\bc{{\bf C}}
\def\br{{\bf R}}
\def\de{\delta}
\def\tr{\triangleleft}
\def\al{\alpha}
\def\la{\langle}
\def\ra{\rangle}
\def\G{\Gamma}
\def\th{\theta}
\def\lm{\ti\lambda}
\def\U{\Upsilon}
\def\jp{{1\over 2}}
\def\js{{1\over 4}}
\def\d{\partial}
\def\ds{\partial_\sigma}
\def\dt{\partial_\tau}
\def\be{\begin{equation}}
\def\ee{\end{equation}}
\def\bea{\begin{eqnarray}}
\def\eea{\end{eqnarray}}
\def\D{{\cal D}}
\def\E{{\cal E}}
\def\G{{\cal G}}
\def\H{{\cal H}}
\def\R{{\cal R}}
\def\T{{\cal T}}
\def\bT{\bar{\cal T}}
\def\F{{\cal F}}
\def\n{{1\over n}}
\def\si{\sigma}
\def\ta{\tau}
\def\ov{\over}
\def\l{\lambda}
\def\L{\Lambda}

\def\pih{\hat{\pi}}
\def\Vt{V^{\ti}}
\def\Ut{U^{\ti}}

\def\e{\varepsilon}
\def\b{\beta}
\def\ga{\gamma}

\begin{titlepage}
\begin{flushright}
{}~
IML 00-28\\
hep-th/0010084
\end{flushright}

\vspace{3cm}
\begin{center}
{\Large \bf Monodromic strings}\\
[50pt]{\small
{\bf C. Klim\v{c}\'{\i}k and S. Parkhomenko}\footnote{Permanent address:
Landau Institute for Theoretical Physics, Chernogolovka, Russia}
\\ ~~\\Institute de math\'ematiques de Luminy,
 \\163, Avenue de Luminy, 13288 Marseille, France}

\vspace{1cm}
\begin{abstract}
We argue that apart from the standard closed and open strings
 one may consider a third possibility that we call
 monodromic strings. The  monodromic  string propagating on a target looks
 like an ordinary open string (a mapping from a segment to the target) but its
space of states is isomorphic to that of a closed string.
 It is shown that the monodromic strings
naturally appear in T-dualizing closed strings moving
on  simply connected targets.
As a nontrivial topology changing example we show that the monodromic strings
on a compact Poisson-Lie group are T-dual to the standard closed strings
propagating on the noncompact dual PL group.

\end{abstract}
\end{center}
\end{titlepage}
\newpage
\noindent 1. The D-brane example  shows \cite{Pol} that one should
not apriori discard "non-standard" string boundary conditions
 from consideration.
In fact, the powerful duality principle require their presence
in string theory. In this paper, we shall introduce other non-standard
string boundary conditions whose existence is required by  Poisson-Lie
T-duality \cite{KS1}. We shall see a posteriori, that this new phenomenon
exists also in the Abelian limit of the PL T-duality where it does not
reduce to the standard momentum-winding change. In fact, it gives
the version of the standard Abelian duality for the simply connected
target. The reader will see that the developed formalism can be viewed
as a sort of continuum version of the discrete orbifold construction
\cite{DHVW}
in the sense the twisted sectors are parametrized by a continuous parameter.
The monodromic PL-T-duality (and its Abelian limit) simply exchanges
the continuous
families of the invariant and twisted sectors.

The problem addressed in this article is the old one: "How to include
zero modes in the Poisson-Lie T-duality story"? A partial answer to this
question was given in \cite{KS3}, where it was shown that the role
of the Abelian momentum-winding lattice  is in general played by the
fundamental group of the underlying Drinfeld double. However, non-Abelian
doubles have small fundamental groups in general, thus the phenomenon
of the momentum-winding exchange does not have much content in the
non-Abelian setting.

We are going to show here, that the non-Abelian
momentum-winding exchange will become a much richer structure if we
release the constraint that the  string should be closed. In fact,
the duality itself will tell us how to "tear up" a closed string.
The disrupture is measured by a certain monodromy and this monodromy
is nothing but  the (non-Abelian) momentum of  the  dual strictly closed
string.

The plan of this paper is as follows. First we shall review the Poisson-Lie
T-duality without the zero modes. In particular, we shall write down
the corresponding  duality invariant action in the Drinfeld double.
Then we shall modify the action on the double by adding a new variable
which will transform into the momentum zero modes of closed strings
if we descend from the double to one of the   Poisson-Lie group targets.
But if we descend to another (dual) target that new variable becomes
a monodromy that measures
how the closed string got torn up.

 We shall finish by a detailed description of this
phenomenon in the context of the Lu-Weinstein-Soibelman \cite{LWS} pair
of the Poisson-Lie groups and also in the context of
the ordinary Abelian T-duality \cite{KY}.

\vskip1pc
\noindent 2. The Poisson-Lie T-duality in its more modern version \cite{KS3}
relates two non-linear $\si$-models living in two different targets $D/G$
and $D/\ti G$. Here $D$ is a Lie group such that i) $dimD=2dimG=2dim\ti G$;
 ii) $G$ and $\ti G$ are  both subgroups of $D$; iii) it exists a symmetric
non-degenerate invariant  bilinear form $(.,.)$ on $\D=Lie(D)$  such that
$(\G,\G)=(\ti\G,\ti\G)=0$, where $\G=Lie(G)$ and $\ti\G=Lie(\ti G)$.
In other words, $\G$ and $\ti\G$ are isotropic subalgebras of $\D$.

Let us state
clearly, that not every group $D$ satisfying the conditions i)-iii) is the
so-called Drinfeld double. For this to be true, it is moreover required
that iv) $\G\cap\ti\G=0$.  However, the modern version
of the Poisson-Lie T-duality does not require iv). In our previous paper
\cite{KP}, we have somewhat abusively called $D$ the "Drinfeld double"
even when the condition iv) was not satisfied.

In order to make this paper technically simpler, we shall study here an older
less general  version of the Poisson-Lie T-duality \cite{KS1,KS2} where
 on the
top of the conditions i)-iii) two more things are  required: the first
is that $D$ is indeed the Drinfeld double (the condition iv) is fulfilled)
and the second is that $D$ is the so-called perfect Drinfeld double,
which means that $D$ can be globally smoothly decomposed as
 $D=G\ti G=\ti GG$.

If the double is perfect, then the Poisson-Lie T-duality exchanges the
targets
$G$ and $\ti G$ because $D/G$ can be identified with $\ti G$ and $D/\ti G$
with $G$. Recall that a Poisson-Lie structure on $G$ is characterized by a
  Poisson
bivector $\alpha\in\Lambda^2TG$ (fulfilling the Jacobi identity)  or,
equivalently, by its right trivialization.
The latter is  a map
$\Pi:G\to\Lambda^2\G$ defined as follows
\be \Pi(g)=R_{g^{-1}*}\alpha_g,\ee
where $R_g$ is the right transport on the group manifold $G$.
It is moreover required  that a  cocycle
condition is fulfilled:
\be \Pi(gh)=\Pi(g) +Ad_g\Pi(h).\ee
Now we can write down
the actions of the corresponding pair of $\si$-models:
\be S_\Pi={1\over 8\pi}\int\la (R+\Pi(g))^{-1},dgg^{-1}\w *dgg^{-1}\ra.\ee
\be S_{\ti\Pi}={1\over 8\pi}
\int\la (R^{-1}+\ti\Pi(\ti g))^{-1},d\ti g\ti g^{-1}\w *d\ti g\ti g^{-1}
\ra.\ee
Here $\ti\Pi(\ti g)$
is the Poisson-Lie structure on the dual group $\ti G$
and $R\in\G\otimes\G$ is some nondegenerate bilinear  form on the dual
space $\G^*$ of the Lie algebra $\G$. Since $\G^*$ can be naturally
identified
with $\ti\G$ via the bilinear form $(.,.)$ on the double $\D$, we may
consider
the inverse  bilinear form $R^{-1}$ as an element
of $\ti\G\otimes \ti\G$. The symbol $\la .,.\ra$ denotes the pairing
between a vector space and its dual.
Finally,
$dgg^{-1}\in T^*\Sigma\otimes \G$ is the pull-back of the right-invariant
 Maurer-Cartan form on $G$ to the
world-sheet $\Sigma$
and $*$ is the Hodge
star on $\Sigma$.

The pair of models (3) and (4)  was first introduced in \cite{KS1}
in somewhat disguised form.
In the  form (3)  and (4), it was rewritten in \cite{KS2}. It is very
important to understand, in which sense these two models are dual to
each other.
 We can study a closed string propagation
on the group $G$ governed by the action (3) and do the same thing for
the target $\ti G$ and the action (4). However, there is no duality in
this case.
In other words,
it is not true that  the Poisson-Lie T-duality relates (3) and (4) as models
of standard closed strings.
The models (3) and (4) become dual only if we remove momentum zero modes from
the closed strings.
In the Abelian context this would mean that the strings may have only
the oscillator
excitations\footnote{As we have already said, there exists a way of
implementing some
(discrete)
momentum modes into the duality story if the groups $G$ and/or $\ti G$
are not
simply connected \cite{KS3}.
If the double $D$ is  perfect then $\pi_1(\ti G)$
parametrize the possible discrete momentum modes and $\pi_1(G)$ parametrizes
the winding modes of the closed strings
moving on $G$. From the point of view of the target $\ti G$,
the momentum-winding
interpretation of the homotopy groups gets exchanged.
This way covers the famous momentum-winding
exchange in the Abelian T-duality context \cite{KY} where $G$ is a
circle group
and $\ti G$ is the dual circle.
In this paper $G$ and $\ti G$ are always simply connected groups;
 we are going to show that the momentum zero modes can be
implemented into the duality story also for this special case at the
price of "tearing up" the closed strings.}.

First of all, let us review \cite{KS1,KS3,KS2} what we mean by the removing
the momentum from the closed string in the non-Abelian context. For this,
 consider
standard closed strings propagating in the simply connected group $G$
according to
the action (3).
Corresponding field equations can be most easily expressed by introducing
certain $1$-form $\lm$ on the world-sheet with values in the dual Lie
algebra $\ti\G$:
\be \lm=-\pi_{\ti\G}Ad_g(R+\Pi(g))^{-1}(\partial_+gg^{-1},.)
 d\xi^++\pi_{\ti\G}Ad_g(R+\Pi(g))^{-1}(.,\partial_-gg^{-1}) d\xi^-.\ee
Here $\xi^{\pm}$ are the usual lightcone variables on the cylinder:
$$\xi^{\pm}=\jp(\tau\pm\si),\quad \tau\in\br,\quad \si\in [0,2\pi];$$
$$\partial_{\pm}=\partial_\tau\pm\partial_\si$$
and $\pi_{\ti\G}$ is a projector to the subalgebra $\ti\G$ with kernel $\G$.
Note that the expression $(R+\Pi(g))^{-1}(\partial_+gg^{-1},.)$ lies
 in $\ti\G$.
We view it as an element of $\D$ and act in the adjoint way by the
element $g\in G$.
The result is projected by $\pi_{\ti\G}$ so that $\lm$ lies in $\ti\G$.
The field equations in terms of $\lm$ have a very simple form:
\be d\lm=\lm\w\lm,\ee
or, in some basis $\ti T_a$ of $\ti\G$:
\be d\lm^a=\jp \ti f^a_{~bc}\lm^b\w\lm^c.\ee
Here $\lm=\lm^a\ti T_a$ and $\ti f^a_{~bc}$ are the structure constants
 of $\ti\G$.
The only nontrivial fact needed for deriving the field equations (6) from
the action (3) is the cocycle condition (2).

We see that every solution $g(\tau,\si)$ of the field equations of the
 model (3)
defines a flat $\ti\G$-valued connection $\lm$. Its monodromy is defined
 by the formula
\be \ti M=P\exp{\int_\gamma \lm(g)},\ee
where $\gamma$ is a curve going around the cylinder. In particular,
we can choose
a curve $\tau=const$. This monodromy is called a noncommutative momentum
\cite{KS1,KS3,KS2}
and its conjugacy class does not depend on time if $g$ is a solution of the
field equations. In particular, if the noncommutative momentum
$\ti M$ is the unit element $\ti e$ of the dual (by assumption also simply
 connected)
group $\ti G$ at some time, then
it will remain $\ti e$ for all times.

 Suppose now that \be \ti M=\ti e\ee
 for some solution $g(\tau,\si)\in G$.
 This means that it exists a single-valued
function $\ti h(\tau,\si)\in\ti G$ on the world-sheet such that
\be \lm(g)=d\ti h\ti h^{-1}.\ee
Consider then the following $D$-valued function $l(\tau,\si)\in D$ on
the worldsheet:
\be l(\tau,\si)=g(\tau,\si)\ti h(\tau,\si).\ee
This mapping can be decomposed as
\be l(\tau,\si)=\ti g(\tau,\si) h(\tau,\si), \quad \ti g\in\ti G,
\quad h\in G,\ee
because of the fact that we have two global decompositions of the double:
$D=G\ti G=\ti G G$.
Then it turns out \cite{KS1} that $\ti g(\tau,\si)$ is a solution of
field equations
of the dual model (4) and a dual $\G$-valued connexion
 $\lambda$ is given by
\be \lambda(\ti g)=dhh^{-1}.\ee
Since the field $h(\tau,\si)$ is evidently single-valued, the dual
 noncommutative
momentum $M$
also satisfies
\be  M=P\exp{\int_\gamma \lambda(\ti g)}=e,\ee
where $e$ is the unit element of $G$.

It is well-known that the phase space of a field theoretical model can
 be viewed
as the space of its classical solutions. Consider the phase
 space $\Upsilon$ of classical
solutions corresponding to closed strings propagating
according to (3) and perform a symplectic reduction
by imposing  the constraint (9) of the unit noncommutative momentum.
We obtain in this way a reduced phase space
$\Upsilon_{\ti e}$. We do the same thing
for the model (4) and we obtain a dual reduced phase space
 $\ti\Upsilon_e$.
Thus both reduced phase spaces
 $\Upsilon_{\ti e}$ and $\ti\U_e$ inherit  symplectic
structures from $\U$ and $\ti\U$, respectively. Moreover, since the unit
noncommutative  momentum
constraints commute with the time evolution, it follows
that $\Upsilon_{\ti e}$ and $\ti\U_e$ inherit
also certain Hamiltonians $H_{\ti e}$ and $\ti H_e$
from the closed string Hamiltonians $H$ and $\ti H$.

 The meaning
of the usual statement
that the models (3) and (4) are related by the Poisson-Lie T-duality is
the following:
There exists a symplectomorphism (preserving the Hamiltonian)
between the dynamical systems $(\U_{\ti e},H_{\ti e})$
and $(\ti\U_{e},\ti H_{e})$.
This symplectomorphism was found in \cite{KS1,KS2}. For its more algebraic
description  see \cite{Sfe}.

There exists
a duality invariant description \cite{KS2}
of the equivalent dynamical systems
 $(\U_{\ti e},H_{\ti e})$ and $(\ti\U_{e},\ti H_{e})$. It turns out
that the phase
space $\U_{\ti e}$ can be identified with the coset $LD/D$
where $LD$ denoted
the loop group of the Drinfeld double $D$.  In other words, $LD$
is the set of smooth
maps from a circle $S^1$ into $D$ equipped with the pointwise multiplication.
The symplectic form $\Omega$ on this  coset can be defined as the
exterior derivative of certain $1$-form $\theta$ on $LD/D$. The latter is
most naturally defined in terms of its integral along an arbitrary curve
$\Gamma$ in the phase space, parametrized by a parameter $\tau$. This curve
can be represented by a certain $D$-valued function $l(\tau,\si)\in D$.
 We define
\be \int_\Gamma\theta={1\over 8\pi}\int (\partial_\si ll^{-1},
\partial_\tau ll^{-1})
+{1\over 48\pi}\int d^{-1}(dll^{-1},[dll^{-1},dll^{-1}]),\ee
where $(.,.)$ is the invariant bilinear form on $\D$ and we recognize
also the
well-known WZW term on the r.h.s..
Note that this definition of $\theta$ is ambiguous since the choice of
the inverse
exterior derivative $d^{-1}$ is too. However, this ambiguity disappears
at the level
of the symplectic form $\Omega=d\theta$.

It may appear that (15) gives the action
of the standard WZW model but this is not quite true, because $\tau$
and $\si$
are not the light cone variables $\xi^\pm$. Nevertheless, the only difference
between the ordinary WZW model and our expression (15) consists in the
{\it names}
of the variables. This means that our expression enjoys
the formal mathematical properties of the standard WZW action.
In particular,
if we replace $l(\tau,\si)$ by $l(\tau,\si)l_0(\tau)$, the integral
$\int\theta$
does not change (the chiral invariance of the WZW model!) hence
the symplectic
form $\Omega$ lives really on the coset $LD/D$ and not on $LD$ itself.

As it is well-known, a first order action of a dynamical system
$(\Omega=d\theta,H_{\ti e})$
is given by
\be S=\int (\theta -H_{\ti e}dt),\ee
where the Hamiltonian $H_{\ti e}$ can be written as follows $\cite{KS2}$:
\be H_{\ti e}=
{1\over 8\pi}\int (\partial_\si ll^{-1},\R \partial_\si ll^{-1}).\ee
Here $\R$ is a linear idempotent self-adjoint map from the Lie algebra $\D$
to $\D$ itself. $\R$ has two equally degenerated eigenvalues $+1$ and $-1$
and the corresponding eigenspaces $R_\pm$ are
\be R_+=Span(t+R(t,.)),t\in \ti\G,\qquad R_-=Span(t-R(.,t)),t\in \ti\G.\ee
Needless to say, $R(.,.)$ is the bilinear form appearing in (3) and (4).
 Putting
(15) and (17) into (16), we obtain the explicit form of the action of the
dynamical system $(\U_{\ti e},H_{\ti e})=(\ti\U_e,\ti H_e)$:
$$ S(l)={1\over 8\pi}\int (\partial_\si ll^{-1},\partial_\tau ll^{-1})
+{1\over 48\pi}\int d^{-1}(dll^{-1},[dll^{-1},dll^{-1}])$$
\be -
{1\over 8\pi}\int (\partial_\si ll^{-1},\R \partial_\si ll^{-1}).\ee
Note that this action is invariant with respect to the gauge transformation
$l(\tau,\si)\to l(\tau,\si)l_0(\tau)$, which means that the model lives
rather on $LD/D$ than on $LD$.

We shall not review the derivation of the constrained models (3)
and (4) from
(19); actually such a derivation is a special case of a more general story
that we are going to present in this paper.

\vskip1pc

\noindent 3. We stress that the duality described so far takes
places between
$(\U_{\ti e},H_{\ti e})$ and $(\ti\U_{e},\ti H_{e})$ and
not between $(\U,H)$
and $(\ti \U, \ti H)$.  In this paper, we want to find a dynamical system
which is dual to $(\U,H)$, in other words: which is dual to the model (3)
describing  closed strings with {\it arbitrary} noncommutative momentum.
The crucial problem to face is the following: if the closed string solution
$g(\tau,\si)$ on $G$
has a non-unit non-commutative momentum, then the configuration (cf. (11))
$l(\tau,\si)\in D$ does not describe a propagation of a
closed string in the double. The reason is that the map $\ti h(\tau,\si)$
defined by (10) is not single-valued on the world-sheet cylinder,
but it developes
some monodromy. If we restrict $\ti h(\tau,\si)$ to the interval
$\si\in[0,2\pi]$,
we obtain a strip propagating in $\ti G$ rather then a cylinder. The same
thing happens for $l(\tau,\si)=g(\tau,\si)\ti h(\tau,\si)$, which does not
correspond to a closed string world-sheet embedded in $D$.
If we project $l(\tau,\si)$ to
$\ti g(\tau,\si)$ according to  (12), the string configuration
$\ti g(\tau,\si)\in \ti G$ will not be closed, i.e.
$\ti g(\ta,\si+2\pi)\neq \ti g(\ta,\si)$.
It is precisely for this reason that the unit non-commutative
momentum constraint
was imposed in \cite{KS1,KS2}.

Our point of view in this paper is very different: We say that
it is not bad
to tear up the closed strings but on the contrary, it is rather
an interesting
thing to do. The point is that the strings in the dual target
get torn up
in a controlled way dictated by duality. In particular, the
duality predicts that
the space of states (=the phase space at the classical level)
of the torn up
string on $\ti G$ must be identical to that of the standard
closed string on $G$.
Thus we obtain a consistent dynamics of open-like strings
whose space of states is that
of the closed string!
We shall  do it in detail for the Lu-Weinstein-Soibelman (LWS)
pair of Poisson-Lie groups.

\vskip1pc

\noindent 4. The LWS double $D$ is simply the
complexification (viewed
as the $real$ group)
$\ti G^{\bf C}$ of a simple compact simply connected and
connected group $\ti G$.
 So, for example,
the LWS
double of $SU(2)$ is  $SL(2,{\bf C})$. The invariant non-degenerate
form $( .,.)$ on the Lie algebra $\D$ of $D$ is given by
\be (x,y)={\rm Im}K(x,y),\ee
or, in other words, it is just the imaginary part of
the Killing-Cartan form $K(.,.)$.
 Since $\ti G$ is the real form of $\ti G^\bc$,
clearly the imaginary part of $K(x,y)$ vanishes if $x,y\in\ti\G$. Hence,
$\ti G$ is indeed isotropically embedded in $\ti G^\bc$.

The dual subgroup  $G$ coincides with
the so called $AN$ group in the Iwasawa decomposition of $\ti G^\bc$:
\be \ti G^\bc=\ti GAN.\ee
For the groups $SL(n,{\bf C})$ the group $AN$ can be identified with
upper triangular matrices of determinant $1$
and with positive real numbers on the diagonal.
In general, the elements of $AN$ can be uniquely represented by means
of the exponential map as follows
\be g={\rm e}^{\phi}{\rm exp}[\Sigma_{\al>0}v_\al E_\al]\equiv
 {\rm e}^{\phi}n.\ee
Here $\al$'s denote the roots of $\ti \G^\bc$, $v_\al$ are complex numbers
and $\phi$ is an Hermitian element\footnote{Recall that the
Hermitian element of  any complex simple Lie algebra $\ti \G^\bc$
is an eigenvector of the
involution which defines the compact real form $\ti\G$;
the corresponding eigenvalue
is $(-1)$ . This involution comes from the group involution
$g\to (g^{-1})^\dagger$.
The anti-Hermitian elements that span
 the compact real form are eigenvectors
of the same involution with the eigenvalue equal to $1$. For elements
of $sl(n,{\rm C})$ Lie algebra, the Hermitian element is
indeed a Hermitian
matrix in the standard sense.}
 of the Cartan subalgebra
of $\ti\G^\bc$. Loosely said, $A$ is the "noncompact part" of the
complex maximal
torus of $\ti G^\bc$. The isotropy of the Lie algebra $\G$ of  $G=AN$
follows from (20); the fact that $\G$ and $\ti \G$ generate together
the Lie algebra $\D$ of the whole double  is evident from (21).

\vskip1pc
\noindent 5. The reason why we have chosen to work with the LWS double
is simple:
both isotropic subgroups $G$ and $\ti G$ are non-Abelian and
one of them ($\ti G$)  is compact and we have
a very good control of the monodromy  valued
in the compact group. Indeed, the non-commutative momentum (8) of
closed string
propagating
on the noncompact group $G=AN$ according to (3)
 takes values in $\ti G$. As we have already remarked, the non-commutative
momenta correspond  to the conjugacy classes in the group $\ti G$.

 It is
well-known \cite{BT}, that if we choose a maximal torus $T$ in $\ti G$, then
every conjugacy class intersects $T$. It is therefore enough
to study  when two elements of the maximal torus lie on the same
conjugacy class.
The maximal torus can be viewed as the quotient of the Cartan
subalgebra $\T$
by the coroot lattice $Q^\vee$ (cf. \cite{BT,FG}). We know that
if two elements
of $\T$ are on the same adjoint orbit of $\ti G$ iff they are
related by the action
of the Weyl group. Thus the fundamental domain of the joint
actions of the Weyl group $W$
and  of the coroot lattice $Q^\vee$  on $\T$ can be identified
with the space
of conjugacy classes of $\ti G$. This fundamental domain is often referred
to as the Weyl alcove.  We shall denote it as $\T_+$.

Now we know that  we need to add degrees of freedom corresponding
to the  non-commutative momenta into the duality invariant
action $S(l)$ in (19). It turns out that the way to do it is very simple;
the new action reads
$$ S(l,\mu)={1\over 8\pi}\int (\partial_\si ll^{-1},\partial_\tau ll^{-1})
+{1\over 48\pi}\int d^{-1}(dll^{-1},[dll^{-1},dll^{-1}])+$$
\be +{1\ov 4\pi}\int (\mu,l^{-1}\d_{\ta}l)-
{1\over 8\pi}\int (\partial_\si ll^{-1}+l\mu l^{-1},
\R (\partial_\si ll^{-1}+l\mu l^{-1})).\ee
Here $l(\tau,\si)\in D$ is a $\si$-periodic $D$-valued maps and
$\mu(\tau)\in\T_+(\subset \ti\G)$.
 This new action is gauge invariant with
respect to the transformations $l(\ta,\si)
\to l(\tau,\si)t(\tau)$ where $t(\tau)\in T$. This means that
the   phase space of this dynamical system  is $(LD/T)\times\T_+$.

We proceed by parametrizing the field configurations $l(\tau,\si)$
according to the $D=G\ti G$ decomposition of the double. In other words,
we parametrize $l$ as $l=g\ti h$, where $g\in AN$ and $\ti h\in \ti G$.
The Polyakov-Wiegmann formula \cite{PW} then says
\be {1\over 8\pi}\int (\partial_\si ll^{-1},\partial_\tau ll^{-1})
+{1\over 48\pi}\int d^{-1}(dll^{-1},[dll^{-1},dll^{-1}])={1\ov 4\pi}
\int(\partial_{\sigma}\ti h\ti h^{-1},g^{-1}\d_{\ta}g)
\ee
and the whole action (23) becomes
\be
S(g,\ti h,\mu)={1\ov 4\pi}
\int\{(\ti \Lambda,g^{-1}(\d_{\ta}-\d_{\si})g)+(g^{-1}\d_{\si}g+
\ti\Lambda,P(g)(
g^{-1}\d_{\si}g+\ti\Lambda))\}.\ee
Here  $P$ is a projector on the subspace $R_-$ with the kernel
$R_+$ (cf. (18)).
Moreover, we denote
\be P(g)=Ad_{g^{-1}}PAd_{g}.\ee
Note that the dependence of the action (25) on $\ti h$ and $\mu$ is
 completely
contained in
\be \ti\Lambda=\partial_{\sigma}\ti h\ti h^{-1}+\ti h\mu \ti h^{-1}.\ee
 Now a  crucial observation is as follows: In distiction
to the case of the
Poisson-Lie T-duality without the zero modes \cite{KS2},
the quantity $\ti\Lambda$
is not constrained  by the unit monodromy constraint. In fact,
 the quantity $\ti\Lambda$ is not
constrained by any constraint whatsoever because , by construction,
it  can have an arbitrary monodromy. This means that we
can regard the action
$S(g,\ti h,\mu)$ as the action $S(g,\ti\Lambda)$
of two unconstrained periodic variables $g,\ti\Lambda$,
 where, moreover, the dependence
on $\ti\Lambda$ is Gaussian. Thus we can solve away
$\ti\Lambda$ from (25) which gives
\be \ti\Lambda =\lm_-(g)-\lm_+(g),\ee
where $\lm_\pm(g)$ were defined in (5). Inserting $\ti\Lambda$
from (28) into
the action $S(g,\Lambda)$, we obtain the action (3):
\be  S_\Pi(g)={1\over 8\pi}\int d\ta d\si
(R+\Pi(g))_{ij}^{-1}[(\d_\ta gg^{-1})^i
(\d_\ta gg^{-1})^j-(\d_\si gg^{-1})^i
(\d_\si gg^{-1})^j].\ee
Here $T_i$ is a basis in $\G=Lie(AN)$ and $\ti T^i$
its dual basis of $\ti\G$ so that
\be (T_i,\ti T^j)=\delta_i^j.\ee
Thus
\be \d_\ta gg^{-1}=(\d_\ta gg^{-1})^i T_i\ee
and \be (R+\Pi(g))^{-1}=(R+\Pi(g))_{ij}^{-1}(\ti T^i\otimes \ti T^j) .\ee
The reader may easily check our derivation
of the model (3) from the duality
invariant action (23) by noting an explicit formula
for the Poisson bivector $\Pi(g)$:
\be \Pi(g)=b(g) a(g)^{-1},\ee
where the matrices $a(g)$ and $b(g)$ are defined as
\be g^{-1}T_i g=a(g)_i^{~j}T_j;\ee
\be g^{-1}\ti T^i g=b(g)^{ij}T_j+d(g)^i_{~j}\ti T^j.\ee
By using the definition (8) of the noncommutative
momentum $\ti M$ and formulas
(27) and (28), we arrive at
\be \ti M=\exp{2\pi\mu}.\ee
Thus the quantity $\mu$ in the duality invariant action (23) becomes
indeed the noncommutative momentum of a
closed string propagating on the target $G=AN$.

\vskip1pc

\noindent 6. So far we have established that the action (3) describing
 (non-constrained) closed strings can be written in
the first-order form (23).
Now we are looking for the dual action. We shall find
out that it is given by a slight
but interesting modification of (4) that corresponds
to a replacement of closed
strings by monodromic strings.

Consider a dual decomposition $l=\ti gh$, where
$\ti g\in \ti G$ and $h\in G(=AN)$.
The Polyakov-Wiegmann formula now reads
\be {1\over 8\pi}\int (\partial_\si ll^{-1},\partial_\tau ll^{-1})
+{1\over 48\pi}\int d^{-1}(dll^{-1},[dll^{-1},dll^{-1}])={1\ov 4\pi}
\int(\partial_{\sigma} h h^{-1},\ti g^{-1}\d_{\ta}\ti g).
\ee
It follows that the action (23) can be rewritten as
$$ S(\ti g, h,\mu)={1\ov 4\pi}\int(\mu,h^{-1}\d_\ta h)+
{1\ov 4\pi}\int (\Lambda,\ti g^{-1}(\d_{\ta}-\d_{\si})\ti g-\mu)+$$
\be +{1\ov 4\pi}\int (\ti g^{-1}\d_{\si}\ti g+\mu+\Lambda,P(\ti g)(
\ti g^{-1}\d_{\si}\ti g+\mu +\Lambda)),\ee
where we have set
\be \Lambda= \d_\si hh^{-1}+h\mu h^{-1}-\mu.\ee
In distinction to the variable $\ti\Lambda$ of the previous case, the
analoguous  quantity $\Lambda$ is now constrained. In order
to understand the nature of this constraint, it is useful to
decompose the variable
$h(\tau,\si)\in AN$ as
\be h(\tau,\si)=e^{\phi(\tau,\si)}n(\tau,\si),
\quad \phi\in Lie(A),\quad n\in N.\ee
Of course, the fields $\phi,n$ are also periodic in $\si$.
The variable $\Lambda$ now becomes
\be \Lambda=\d_\si\phi+e^{\phi}(\d_\si nn^{-1} +n\mu n^{-1}-\mu)
e^{-\phi}\equiv
\Lambda_A+\Lambda_N.\ee
We observe immediately, that $\Lambda_A$ is in $Lie(A)$
and $\Lambda_N$ is in $Lie(N)$.

By a straightforward  study of adjoint orbits of Borel
subgroups of $\ti G^\bc$,
we arrive at conclusion that $\Lambda_N$ is not constrained. In other
words,  by varying $h(\ta,\si)$ and $\mu(t)$,  one completely
sweeps the space of all posible $\Lambda_N(\ta,\si)$. On the other hand,
$\Lambda_A(\ta,\si)$ is clearly
constrained, it misses the zero mode in the Fourier series
in the variable $\si$.
Such a constraint  can be easily taken into account by adding
to the action (38)
a Lagrange multiplier term $\int (\nu(\tau),\Lambda_A(\tau,\si))$,
 where $\nu(\tau)$
is in the Cartan subalgebra $\T(\subset \ti\G)$. Thus the action (38) gets
transformed into the following equivalent one:
$$ S(\ti g,\Lambda,\nu,\mu,\phi_0)=
-{1\ov 4\pi}\int \{(\d_\ta\mu,\phi_0)-(\Lambda,\nu)\}+
{1\ov 4\pi}\int (\Lambda,\ti g^{-1}(\d_{\ta}-\d_{\si})\ti g-\mu)+$$
\be +{1\ov 4\pi}\int (\ti g^{-1}\d_{\si}\ti g+\mu+\Lambda,P(\ti g)(
\ti g^{-1}\d_{\si}\ti g+\mu +\Lambda)),\ee
where
\be \phi_0(\tau)=\int d\si \phi(\ta,\si).\ee
Recall that we wish to solve away the  variables $\Lambda$ and
 $\phi_0$ (related
to the variable $h$ in the decomposition $l=\ti gh$). This can
be down easily,
since after the introducing the Lagrange multiplier $\nu$,
those variables $\Lambda,\phi_0$ are unconstrained in (42). The resulting
action is
$$ S_{\ti\Pi}(\ti g,\nu,\mu_0)
 ={1\over 8\pi}\int d\ta d\si (R^{-1}+\ti\Pi(\ti g))_{ij}^{-1}$$
\be [(\d_\ta \ti g\ti g^{-1}+\ti g
\nu \ti g^{-1})^i
(\d_\ta \ti g\ti g^{-1}+\ti g
\nu \ti g^{-1})^j-(\d_\si \ti g\ti g^{-1}+\ti g
\mu_0 \ti g^{-1})^i
(\d_\si \ti g\ti g^{-1}+\ti g
\mu_0 \ti g^{-1})^j
].\ee
Note that in this final expression (44) for the dual action,
we use the symbol
$\mu_0$ instead of $\mu$. It is because $\mu_0$ is a constant
not depending on
the time $\ta$. This is dictated by integrating away
the Lagrange multiplier $\phi_0$ in the
action (42). On the other hand, $\nu$ still depends on $\ta$.

We would like to interpret our result. First of all, note that
$\ti g(\ta,\si)$'s
are  $\ti G$-valued functions periodic in $\si$, so we could view (44)
as a  $\mu_0$-depending dynamical system describing closed
string configurations interacting with some particle-like degrees
of freedom
$\nu$. Such a theory would  not be anymore a $\si$-model in the
standard sense
of this word. For example,  $\ti g$-depending
terms not containing derivatives of $\ti g$ also appear in the
action which would
mean that our duality transformation has generated a tachyon potential.

We believe that the correct interpretation is the following:
Introduce a new
field variable
\be \ti m(\ta,\si)=\ti g(\ta,\si)e^{\mu_0\si}.\ee
Such a configuration is referred to as the monodromic string for  obvious
reasons. Its image in the target $\ti G$ looks like an open string.
The action
(44) can be then rewritten as
$$ S_{\ti\Pi}(\ti m,\nu)
 ={1\over 8\pi}\int d\ta d\si (R^{-1}+\ti\Pi(\ti m))_{ij}^{-1}$$
\be [(\d_\ta \ti m\ti m^{-1}+\ti m
\nu \ti m^{-1})^i
(\d_\ta \ti m\ti m^{-1}+\ti m
\nu \ti m^{-1})^j-(\d_\si \ti m\ti m^{-1})^i
(\d_\si \ti m\ti m^{-1})^j
].\ee
The reader may ask why we are allowed to  replace $\ti\Pi(\ti g)$ by
 $\ti\Pi(\ti m)$.
We have from (2)
\be  \ti\Pi(\ti m)=\ti\Pi(\ti ge^{\mu_0\si})=\ti\Pi(\ti g)+
Ad_{\ti g}\ti\Pi(e^{\mu_0\si})=
\ti\Pi(\ti g).\ee
The last equality follows from the fact that the Lu-Weinstein-Soibelman
Poisson-Lie structure $\ti\Pi(\ti g)$ on a
 compact simple connected and simply connected group
$\ti G$ always vanishes on the maximal torus $T\subset \ti G$ \cite{FR}.

We can also naturally interpret the variable $\nu(\tau)\in\T$. In fact
 it is
the gauge field. By using again the fact that $\ti\Pi$ vanishes
 on the maximal torus,
we observe that the model (4) has a global symmetry
$\ti g\to\ti gt$, where
$t\in T$. This global symmetry  is present for the closed but
also for the monodromic
string. Its gauging  amounts for introducing the gauge fields
$\nu$ into the action (4). This is precisely the action (46).
The gauge tranformation
reads
\be \ti m\to\ti mt(\ta),\quad \nu\to \nu-t^{-1}\d_\ta t.\ee

Summarizing, we have obtained the following picture: The closed string
model (3) on $G=AN$ is dual to the monodromic string model (4)
on the compact
group $\ti G$, where the
maximal torus momentum zero modes are gauged away. In some sense we may say,
that the duality
requires adding the monodromy zero modes to (4) but at the same time
removing  maximal torus momentum zero modes.
\vskip1pc

\noindent 7. Abelian example. It turns out that one has a
full control of the
monodromy also in an almost trivial but instructive example
where the Drinfeld
double $D$ is just a plane $\br^2$ viewed as the additive
Abelian Lie group.
Its Lie algebra is again $\br^2$ and the exponential map
is just the identity
map. The invariant
bilinear form is defined as
\be ((x_1,y_1),(x_2,y_2))=x_1y_2+y_1x_2.\ee
It is  symmetric and non-degenerate. Then we define
\be\G=\{(x,0);x\in \br\},\quad \ti\G=\{(0,\ti x); \ti x\in\br\}.\ee
Both subalgebras $\G$,$\ti\G$ are isotropic and $\G\cap\ti\G=0$.
The double $D$ is moreover clearly perfect, i.e. $D=G+\ti G$.

We can immediately write down the action (23) for our
Abelian double, where
$l(\ta,\si)=x(\ta,\si)+\ti x(\ta,\si)$, $x(\ta,\si)\in G$,
$\ti x(\ta,\si)\in\ti G$ and
 $\mu(\ta)\in\ti\G$. The action reads
$$ 4\pi S(x,\ti x,\mu)=\int \d_\si \ti x\d_\ta x+\int \mu\d_\ta x
-\jp R\int (\d_\si \ti x+\mu)^2-\jp R^{-1}\int (\d_\si x)^2=$$
\be=\int \ti\Lambda \d_\ta x -\jp R\int
\ti\Lambda^2-\jp R^{-1}\int (\d_\si x)^2,\ee
where $R$ is a positive real number. We observe that the quantity
 $\ti\Lambda=\d_\si\ti x+\mu$ is unconstrained, hence we
may solve it away to obtain
\be S(x)={1\ov 8\pi R}\int [(\d_\ta x)^2-(\d_\si x)^2].\ee
The same action (51) can be rewritten from the dual point of view as
$$ 4\pi S(x,\ti x,\mu)=\int \d_\si  x\d_\ta \ti x-\int \d_\ta\mu x_0
-\jp R\int (\d_\si \ti x+\mu)^2-\jp R^{-1}\int (\d_\si x)^2,$$
where $x_0(\tau)$ is the zero Fourier component of the field $x(\ta,\si)$.

By repeating the same procedure as in the general case, we arrive at the
following dual action
\be \ti S(\ti x,\nu,\mu_0)=
{R\ov 8\pi}\int [(\d_\ta \ti x+\nu)^2-(\d_\si \ti x+\mu_0)^2],\ee
where the field $\ti x$ is still periodic. We can decompose $\ti x$
as
\be \ti x=\ti x_0+ \ti x_{osc},\ee
where $\ti x_0$ denotes the zero Fourier component and
$\ti x_{osc}$ the rest.
Then define
\be \ti x_{mon}=\ti x_{osc}+\mu_0\si.\ee
Finally observe that the zero modes $\ti x_0$ can be
absorbed into $\nu$ by setting
\be \nu'=\nu+\d_\ta \ti x_0.\ee The resulting action reads
\be \ti S(\ti x_{mon},\nu')={R\ov 8\pi}\int [(\d_\ta \ti x_{mon}+
\nu')^2-(\d_\si \ti x_{mon})^2].\ee
The field $\nu'$ can be solved away to yield our final dual result
\be \ti S(\ti x_{mon})={R\ov 8\pi}\int
[(\d_\ta \ti x_{mon})^2-(\d_\si \ti x_{mon})^2].\ee
It is also instructive to calculate the noncommutative momentum $\ti M$ of
the standard closed strings living on $G$. For this, we have to solve
the equations of motion of the model (51). The general solution
$x_{sol}$ reads
\be x_{sol}(\ta,\si)= x_0+p\tau +osc_L(\ta-\si)+osc_R(\ta+\si).\ee
The quantity $\lm$ then reads
\be \lm=\d_- x_{sol} d\xi^- -\d_+x_{sol}d\xi^+\ee
and the momentum $\ti M$
\be \ti M=-{p\ov R}.\ee
We observe that in the Abelian case, the noncommutative momentum becomes
the standard momentum of the  closed string.

\vskip1pc
\noindent 8. Dressing cosets. There exists the generalization of the
Poisson-Lie T-duality \cite{KS4} relating models living on the double cosets
$F\backslash D/G$ and $F\backslash D/\ti G$, where $F$ is certain isotropic
subgroup of $D$. We are going to show now that the story of the monodromic
strings generalizes to this case. Recall first the basic ingredients of the
dressing coset construction.

Consider now an $n$-dimensional linear subspace $\R_+\subset\D$ such
that
it intersects with  its orthogonal
complement $\R_-$ in an isotropic Lie algebra $\F$, i.e.
\be \R_+\cap\R_-=\F; \qquad [\F,\F]\subset\F.\ee
Moreover, both $\R_+$ and $\R_-$ should be invariant subspaces with respect
to the adjoint action of $\F$:
\be [\F,\R_+]\subset\R_+,\qquad  [\F,\R_-]\subset\R_-.\ee
It was shown in \cite{KS4} that all these data define a
 pair of dual non-linear
$\si$-models living respectively, on the targets $F\backslash D/G$ and
$F\backslash D/\ti G$. Their common dynamics is encoded in the first order
Hamiltonian action \cite{KS4} which can be obtained by setting $\mu=0$
in the following more general action principle:
$$ S={1\over 8\pi}\int {\rm d}\si{\rm d}\tau ( \ds ll^{-1},\dt ll^{-1})+
{1\over 48\pi}\int d^{-1}( dll^{-1},[dll^{-1},dll^{-1}]) $$
$$+{1\over 4\pi}\int (\mu, l^{-1}\dt l) -
{1\over 8\pi}\int {\rm d}\si{\rm d}\tau
 \{( (\ds ll^{-1}+l\mu l^{-1})_0,(\ds ll^{-1}+l\mu l^{-1})_0)
$$\be -( (\ds ll^{-1}+l\mu l^{-1})_1,(\ds ll^{-1}+l\mu l^{-1})_1)\}.\ee
The reader has certainly understood that the action with the incorporated
variable $\mu(\tau)\in\T_+$ is   the generalization
of the dressing coset story to the
monodromic case.
Here as before  $l(\si,\tau)=l(\si+2\pi,\tau)$ is a
 mapping from a cylindrical worldsheet
into the group manifold $D$  and $\ds ll^{-1}+l\mu l^{-1}$
is constrained to lie in $\F^\perp$:
\be \ds ll^{-1}+l\mu l^{-1}\in\F^\perp.\ee
Note that due to the non-degeneracy of the form $( .,.)$ we have
\be\F^\perp={\rm Span}(\R_+,\R_-).\ee
We should also explain the meaning of the subscripts $0$ and $1$ in (64).
We can write arbitrary element
 $x\in\F^\perp$ as
\be x=x_0+x_1, \qquad x_0\in\R_+,\quad x_1\in\R_-.\ee
Of course, this decomposition is not unique, because the linear spaces
$\R_+$ and $\R_-$ intersect at $\F$. The decomposition
$ x=x_0' +x_1'$, where $x_0'=x_0+\phi$ and
$x_1'=x_1-\phi$,  is equally good for an arbitrary $\phi\in\F$. However,
due to the fact that $(\F,\R_+)=(\F,\R_-)=0$, the action (64)
does not depend on this decomposition.

The action (64) possesses the following   gauge symmetry
$l\to fl, f(\si,\tau)\in F$ which explains why we take the left coset
 $F\backslash D/G$.

The way how to obtain the dual pair of the $\si$-models from the action (64)
(for $\mu=0$)
was described in \cite{KS4}. The argument  for the nonvanishing
 $\mu$ goes in the
similar way giving as the pair of the dual $\si$-models nothing but the
actions (29) and (46). The reader may ask what is then the difference with
the previous case (with no constraint of the form (65)).
The answer is simple,
if the condition (62) holds for $\R_+$, then the model (46)
 does not live on the
target $D/G$, because it developes additional gauge symmetry coming from
$l\to fl, f(\si,\tau)\in F$. Due to this gauge symmetry
 the target space of the
model (46) is in fact $F\backslash D/G$ instead of $D/G$.
 Reasoning in the same way
gives that the dual model (29) developes also the additional
gauge symmetry and it
lives on $F\backslash D/\ti G$.
\vskip1pc
\noindent 9. Conclusions and outlook: The Poisson-Lie
T-duality story \cite{KS1} is
 the generalization of the traditional non-Abelian T-duality \cite{OQ}.
Apart from the papers already cited in the text, we may
mention several other
works who have developed various aspects of the construction \cite{Ost}.
A quantum version of the PL T-duality is in preliminary stage
\cite{AKT}. We believe that
our present article will facilitate the work in the
quantum direction since
we did get finally rid of the highly nonlinear monodromy constraints.
Thus we hope that the quantum monodromic strings may
become a part of the standard
superstring theories.
\vskip1pc
\noindent 10. Acknowledgement: S. P. acknowledges a support from grants
RBRF-99-02016687, RBRF-96-15-96821, INTAS-OPEN-97-1312 and  RP1-2254.


\begin{thebibliography}{19}

\bibitem{Pol}{J. Polchinski,  {\it Phys.Rev.Lett.}
{\bf 75} (1995) 4724}
\bibitem{KS1}{C. Klim\v c\'\i k and P. \v Severa,
{\it Phys. Lett.} {\bf B351}
 (1995) 455; C. Klim\v c\'\i k, {\it Nucl. Phys. (Proc. Suppl.)} {\bf
B46} (1996) 116; P. \v Severa,
{\it Minim\'alne plochy a dualita}, Diploma thesis, 1995, in Slovak}
\bibitem{DHVW}{L. Dixon, J.A. Harvey, C. Vafa and E. Witten,
 {\it Nucl. Phys.}
{\bf B261} (1985) 678}
\bibitem{KS3}{C. Klim\v c\'\i k and P. \v Severa,
 {\it Phys. Lett.} {\bf B383} (1996) 281}
\bibitem{LWS}{J.-H. Lu and A. Weinstein, {J. Diff. Geom.}
 {\bf 31} (1990) 510;
Ya.S. Soibelman, {\it Algebra Analiz} {\bf 2}
 (1990) 190}
\bibitem{KY}{K. Kikkawa and M. Yamasaki, {\it Phys. Lett.} {\bf B149}
 (1984) 357;
N. Sakai and I. Senda, {\it Prog. Theor. Phys.} {\bf 75} (1986) 692}

\bibitem{KP} {C. Klim\v c\'\i k and S. Parkhomenko,
{\it Phys. Lett.} {\bf B463}
(1999) 195}
\bibitem{KS2}{C. Klim\v c\'\i k and P. \v Severa,
 {\it Phys. Lett.} {\bf B372} (1996) 65}

\bibitem{Sfe}{K. Sfetsos ,{\it Nucl. Phys.} {\bf B517} (1988) 549}

\bibitem{BT}{M. Blau and G. Thompson, {\it Lectures on 2d gauge theories:
Topological aspects
and path integral techniques}, in {\it Proceedings of the 1993
Trieste Summer School
on High Energy Physics and Cosmology} (eds. E. Gava et al.),
World Scientific, Singapore (1994)
175, hep-th/9310144}
\bibitem{FG}{F. Falceto and K. Gaw\c edzki, {\it J.Geom.Phys.}
 {\bf 11} (1993)
251}
\bibitem{PW}{A.Polyakov and P.B.Wiegmann, {\it Phys. Lett.}
{\bf B311} (1983) 549}
\bibitem{FR}{H. Flaschka and T. Ratiu, {\it A convexity theorem for
Poisson actions of compact Lie
groups}, preprint IHES/M/95/24 (1995)}
\bibitem{OQ}{X. de la Ossa and F. Quevedo, {\it Nucl. Phys.} {\bf B403}
 (1993) 377; B.E. Fridling and A. Jevicki, {\it Phys. Lett.} {\bf B134}
(1984) 70;
E.S. Fradkin and A.A. Tseytlin, {\it Ann. Phys.} {\bf 162} (1985) 31}
\bibitem{Ost}{K. Sfetsos,
{\it Nucl. Phys. (Proc. Suppl.)} {\bf
56B} (1997) 302;
{\it Phys. Rev.} {\bf D57} (1998) 3585; M.A. Lledo and V.S. Varadarajan,
{\it Lett. Math. Phys.}{\bf 45}
(1998) 247; S. Parkhomenko, {\it Nucl.Phys.} {\bf B510} (1998) 623;
 A. Stern, {\it Phys. Lett.} {\bf B450} (1999) 141;
{\it Nucl.Phys.} {\bf B557} (1999) 459;
M.A. Jafarizadeh and A. Rezaei-Aghdam, {\it Phys.Lett.}
{\bf B45} (1999) 477;
S. Majid and E.J. Beggs, {\it Poisson-Lie T-duality for
quasitriangular Lie bialgebras}, Cambridge preprint, math.QA/9906040;
O. Alvarez, {\it Nucl.Phys.}{\bf B584} (2000) 659; 682;
F. Assaoui, N. Benhamou and T. Lhallabi,
{\it Poisson-Lie T-duality in supersymmetric WZNW model}, hep-th/0009024 }
\bibitem{AKT}{A. Alekseev, C. Klim\v c\'\i k and
 A. Tseytlin, {\it Nucl.Phys.}
{\bf B458} (1996) 430; E. Tyurin and R. von Unge, {\it Phys.Lett.}
{\bf B382} (1996) 233; P. \v Severa,{\it Quantum Kramers-Wannier duality
and its topology}, hep-th/9803201; K. Sfetsos,
{\it Phys. Lett.} {\bf B432} (1998) 365;
L.K. Balazs, J. Balog, P. Forgacs, N. Mohammedi, L. Palla and
J.Schnittger,
{\it Nucl. Phys.} {\bf B535} (1998) 461;
S. Parkhomenko, {\it J.Exp.Theor.Phys.}
{\bf 89} (1999) 5}
\bibitem{KS4}{C. Klim\v c\'\i k and P. \v Severa,
 {\it Phys. Lett.} {\bf B381} (1996) 56}


\end{thebibliography}
\end{document}